# Main Text: Doping as a tuning mechanism for magneto-thermoelectric effects to improve $zT$ in polycrystalline NbP

Eleanor F. Scott[1], Katherine A. Schlaak[1,2], Poulomi Chakraborty[3], Chenguang Fu[4,5], Satya N. Guin[5,6], Safa Khodabakhsh[1], Ashley E. Paz y Puente[1], Claudia Felser[5], Brian Skinner[3], Sarah J. Watzman[1*]

1. Department of Mechanical and Materials Engineering, University of Cincinnati, Cincinnati, OH 45221
2. Department of Physics, University of Cincinnati, Cincinnati, OH 45221
3. Department of Physics, The Ohio State University, Columbus, OH 43210
4. Department of Materials Science and Engineering, Zhejiang University, Hangzhou, China 310027
5. Max Planck Institute for Chemical Physics of Solids, Dresden, Germany 01187
6. Department of Chemistry, Birla Institute of Technology and Science, Pilani - Hyderabad Campus, Hyderabad, India 500078
*Please direct correspondence to watzmasj@ucmail.uc.edu

**Abstract**

Weyl semimetals combine topological and semimetallic effects, making them candidates for interesting and effective thermoelectric transport properties. Here, we present experimental results on polycrystalline NbP, demonstrating the simultaneous existence of a large Nernst effect and a large magneto-Seebeck effect, which is typically not observed in a single material at the same temperature. We compare transport results from two polycrystalline samples of NbP with previously published work, observing a shift in the temperature at which the maximum Nernst and magneto-Seebeck thermopowers occur, while still maintaining thermopowers of similar magnitude. Theoretical modeling shows how doping strongly alters both the Seebeck and Nernst magneto-thermopowers by shifting the temperature-dependent chemical potential, and the corresponding calculations provide a consistent interpretation of our results. Thus, we offer doping as a tuning mechanism for shifting magneto-thermoelectric effects to temperatures appropriate for device applications, improving $zT$ at desirable operating temperatures. Furthermore, the simultaneous presence of both a large Nernst and magneto-Seebeck thermopower is uncommon and offers unique device advantages if the thermopowers are used additively. Here, we also propose a unique thermoelectric device which would collectively harness the large Nernst and magneto-Seebeck thermopowers to greatly enhance the output and *zT* of conventional thermoelectric devices.

**I. Introduction**

Topological materials have gained much interest in solid-state physics fields since their discovery, with recent work showing that Dirac semimetals and Weyl semimetals (WSMs) are excellent candidates for both longitudinal and transverse magneto-thermoelectric transport applications, including the conversion of waste heat to useful electric power [1,2,3,4,5,6,7,8,9, 10,11,12,13]. Because WSMs are two-carrier systems, it was predicted that these materials would possess a very large Nernst effect and a minimal Seebeck effect. However, WSMs were also theoretically predicted to possess thermomagnetic transport signatures beyond that found in classical semimetals due to their ultra-high mobility charge carriers found in Dirac bands and their bulk Weyl nodes [14,15].

The Seebeck effect is the conventional, longitudinal thermoelectric (TE) effect, while the Nernst effect is its transverse counterpart. In a longitudinal geometry, charge carriers condense on the cold side of the material, making it such that the contributions of oppositely charged carriers effectively cancel each other. Because of this, conventional TE materials are typically single-carrier systems. In contrast, the Nernst effect is produced by applying a magnetic field perpendicular to the applied temperature gradient, producing a Lorentz force and thus a resultant electric field in the mutually orthogonal direction. This magnetic field causes charge separation in the direction of the resulting electric field, meaning the contributions of oppositely-signed charge carriers effectively add together. Therefore, two-carrier systems are expected to possess a large Nernst effect [16]. Although this transverse geometry requires a magnetic field, the temperature gradient and resulting electric field are ultimately decoupled, which gives geometrical freedom in engineering devices [17]. Furthermore, WSMs are predicted to simultaneously exhibit a large magneto-Seebeck effect due to their gapless band structure not seen in traditional semimetals [18,19]. Because both a longitudinal and transverse TE voltage can be produced in the presence of a magnetic field, this class of material presents a unique opportunity to utilize magneto-thermoelectric effects.

Recent experimental results [1,2,20] have led this research to focus on the WSM NbP, a Type I WSM which breaks inversion symmetry [21]. Previous work in single-crystalline NbP observed an unprecedentedly large Nernst thermopower, exceeding 800 $\mu$V K$^{-1}$ at 109 K and 9 T [1]. Initial work in polycrystalline NbP found that a large Nernst thermopower was maintained, ~90 $\mu$V K$^{-1}$ at 136 K and 9 T, and a competitive Nernst power factor was observed over a broad temperature range of approximately 100 K to 200 K [2]. This large Nernst thermopower and Nernst power factor were also generally maintained as Nb vacancies were introduced in off-stoichiometric polycrystalline NbP samples [20]. While these results are motivating, they currently show that TE properties are only maximized at high magnetic fields and cryogenic temperatures with moderate TE figures of merit (*zT*). Here, we focus on polycrystalline, rather than single-crystalline, NbP since polycrystalline samples are more durable, less expensive, and easier to synthesize, making them better-suited for TE device applications, especially as previous results demonstrate a large Nernst power factor in polycrystalline NbP [2,20]. We demonstrate that NbP has not only a large Nernst effect but also a comparably large magneto-Seebeck effect, an unusual feature to find coexisting in semimetallic materials. With the groundwork of previous studies on NbP in mind, this work focuses on what and how material properties can be altered determining appropriate tuning mechanisms to maximize TE effects, and thus *zT*, in NbP. We offer this research as evidence for doping, and thus altering the temperature-dependent chemical potential, as a tuning mechanism for increasing magneto-thermoelectric transport and the temperature range over which it is maximized in NbP. Additionally, we propose a new TE device utilizing a simultaneously large magneto-Seebeck and Nernst effect, which offers the potential of increasing device *zT* by nearly four-fold.

## II. Experimental Methods

Polycrystalline NbP samples were synthesized via direct reaction using powders of niobium and phosphorus. NbP powder was formed into a solid ingot using spark plasma sintering [2], with approximate density of 90%. Two separate bar-shaped samples were cut from the sintered ingot to be used in transport measurements. Two bars of NbP were selected for analysis: Sample 1 was left unannealed with an average grain size of 2.18 μm, and Sample 2 was annealed at 1000 °C for 1 week which resulted in an increased average grain size of 3.08 μm. Sample 2 was wrapped in Nb foil with excess NbP powder then encapsulated in a quartz tube which had been purged with argon and then sealed in a vacuum. The average grain size of both samples was found using electron backscattering diffraction (EBSD) on a scanning electron microscope (SEM). The stoichiometry of both samples was estimated using EDS. This revealed that Sample 2 lost phosphorus during the annealing process leaving it as Nb-rich. This was also confirmed in the calculated charge carrier concentration of both samples, explained further in the Supplementary Materials [22]. The mean free path of NbP was estimated to be ≈ 30 nm, which is much smaller than the average grain size of both samples, therefore the transport differences between Sample 1 and Sample 2 are attributed primarily to the differences in stoichiometry and not grain size.

All transport measurements were completed on a 9 T Quantum Design DynaCool Physical Property Measurement System (PPMS). The thermal conductivity, electrical resistivity, Seebeck effect, and Nernst effect were measured using the Thermal Transport Option (TTO) in the conventional one-heater two-thermometer configuration, with the exception of Sample 2's 9 T electrical resistivity, which was measured on the Electrical Transport Option (ETO). Electrical resistivity and thermal conductivity were measured using MultiVu, Quantum Design's proprietary software for the PPMS. Both the Seebeck effect and Nernst effect were characterized using a home-built electrical breakout system with external electronics and customized controls code written in LabVIEW. The magneto-Seebeck effect was characterized in both steady-state and sweeping magnetic fields, while the Nernst effect was characterized in a sweeping magnetic field. The Hall effect was measured with the PPMS's ETO using the van der Pauw method. Thermoelectric transport data for both the Seebeck and Nernst effects presented in this work was measured isothermally, as opposed to using a conventional adiabatic sample mount where heat flow is unrestricted in the sample and the temperature gradient is not necessarily parallel to only the direction in which it is applied. Isothermal preparation was necessary due to the large thermal Hall effect in NbP, which causes a parasitic temperature gradient to occur when a magnetic field is applied. Further information on isothermal sample preparation for TE measurements can be found in the Supplementary Materials [22].

### III. Experimental Results

The data presented here follows conventional labeling in order to define the transport geometry using the subscript indices "*abc*" where *a* is the direction of the applied flux, *b* is the direction of the measured field, and *c* is the direction of the externally applied magnetic field (if present) [17]. Any value with the *c* index dropped indicates that no magnetic field was applied.

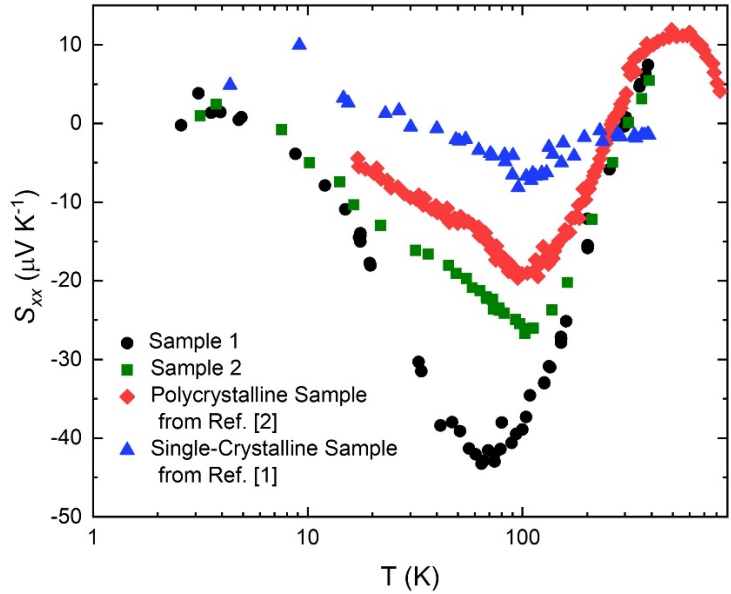

*Figure 1. Comparison of the temperature dependence of the thermopower, $S_{xx}$, in NbP*

The temperature dependence of the Seebeck effect was measured and reported as a thermopower, $S_{xx}$, in Figure 1. All data presented was measured isothermally, except for the polycrystalline sample from Ref. [2], which was measured adiabatically. The Seebeck effect of NbP Sample 1 resulted in a maximum magnitude of thermopower of 43.3 $\mu$V K$^{-1}$ at 64.5 K, and Sample 2 resulted in a maximum magnitude of thermopower of 26.7 $\mu$V K$^{-1}$ at 102.5 K. In comparison, the polycrystalline sample studied in Ref. [2] reached a maximum magnitude of thermopower of ~20 $\mu$V K$^{-1}$ near 95 K, and the single-crystalline sample from Ref. [1] had a maximum magnitude of thermopower of only ~8 $\mu$V K$^{-1}$ near 100 K. All three NbP polycrystalline samples have a thermopower significantly exceeding that of the single-crystalline sample. Furthermore, the temperature at which the maximum magnitude of thermopower occurs changes from sample-to-sample, with the most significant difference being that between Samples 1 and 2.

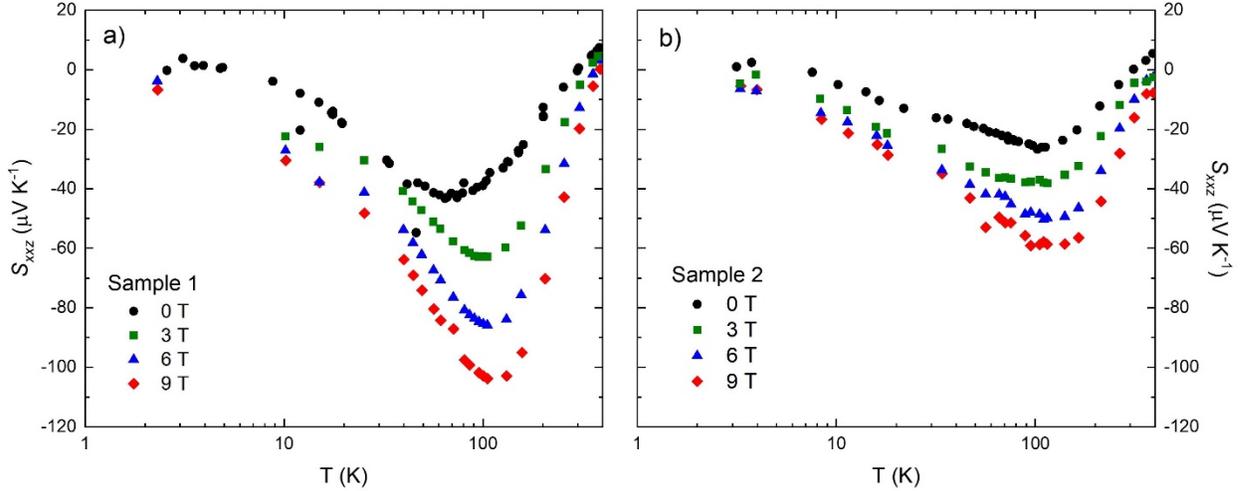

*Figure 2. Temperature- dependence of the magneto-Seebeck thermopower, $S_{xxz}$, at discrete magnetic fields in polycrystalline NbP for (a) Sample 1 and (b) Sample 2*

A large magneto-Seebeck effect, $S_{xxz}$, was observed in both Samples 1 and 2 with the magnetic field applied perpendicular to the directions of both the applied heat flux and measured electric field; this is shown as a function of temperature at discrete magnetic fields in Figure 2. $S_{xxz}$ increased in magnitude as a function of magnetic field up to 9 T, more than doubling the maximum magnitude of thermopower in comparison to the 0 T results. A maximum magnitude of thermopower of 103.8 $\mu$V K$^{-1}$ at 105.4 K, 9 T was observed in Sample 1, and a maximum magnitude of thermopower of 59.1 $\mu$V K$^{-1}$ at 94.77 K, 9 T was observed in Sample 2. Once again, the maximum values of $S_{xxz}$ differ between the samples, and the temperature at which the maximum occurs increases with an increase in externally applied magnetic field for Sample 1. Most notable, though, is the presence of the large isothermal magneto-Seebeck effect in the polycrystalline NbP samples studied in this work. In comparison, previous work in single-crystalline NbP observed no measurable isothermal magneto-Seebeck effect [1], although an adiabatic magneto-Seebeck effect was observed and attributed to parasitic contributions from the Nernst effect into the magneto-Seebeck effect as seen in the Supplementary Materials of Ref [1]; only adiabatic results for the magneto-Seebeck effect have been previously reported in polycrystalline NbP [2].

The Nernst thermopower, $S_{xyz}$, in both Samples 1 and 2, shown as a function of magnetic field at select temperature in Figure 3, was found to be on the same order of magnitude as the magneto-Seebeck thermopower. $S_{xyz}$ in Samples 1 and 2 is comparable to that of polycrystalline NbP in Ref. [2], demonstrating that, although $S_{xyz}$ is large in polycrystalline NbP, it is reduced by at least an order of magnitude from the huge Nernst thermopower observed in single-crystalline NbP [1]. Sample 1 has a maximum Nernst thermopower of 51.6 $\mu$V K$^{-1}$ at 159.1 K, 9 T, while Sample 2 has a maximum Nernst thermopower of 99.7 $\mu$V K$^{-1}$, at 163.8 K, 9 T. Both the magnitudes of the Nernst thermopower and the temperature at which they occur differ between the two samples. The Nernst thermopower as a function of magnetic field also exhibits two distinct slopes, a high field region between |3 T| and |9 T| and a low field region below |2 T|, behavior which is consistent with previously work in NbP [1,2].

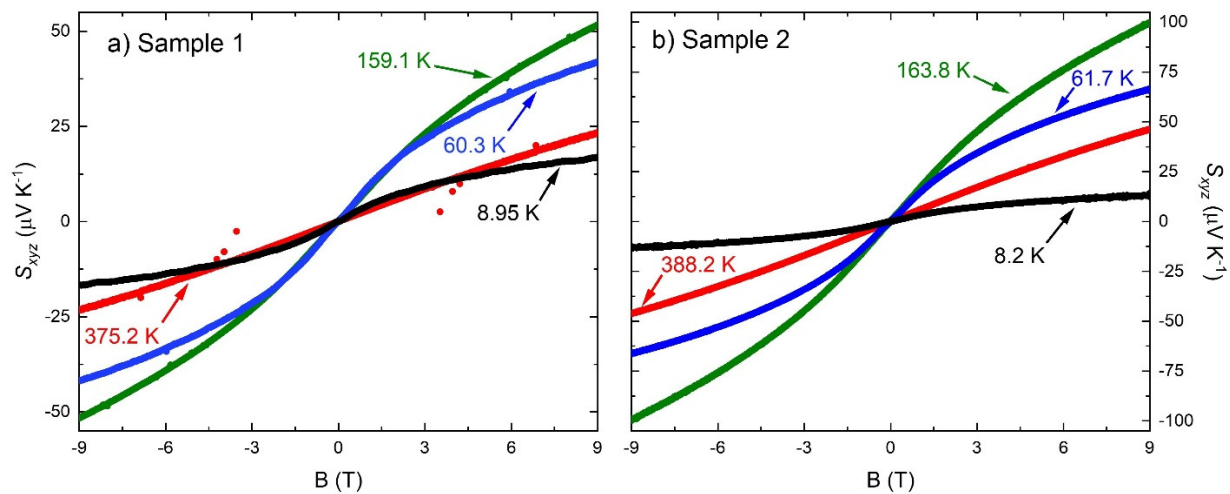

*Figure 3. Magnetic field-dependence of the Nernst thermopower, $S_{xyz}$, at discrete temperatures in polycrystalline NbP for (a) Sample 1 and (b) Sample 2*

Temperature-dependence of the electrical resistivity for Samples 1 and 2 is shown in Figure 4. In the polycrystalline samples, the electrical resistivity is increased from that of the single crystal [1], although the electrical resistivity is still comparable in order of magnitude to that of a conventional metal. The temperature-dependence of thermal conductivity is shown in Figure 5, indicating a strong reduction in thermal conductivity of polycrystalline NbP as compared to single-crystalline NbP. This, combined with the relatively low electrical resistivity, ultimately lends itself to being advantageous for using polycrystalline NbP in thermoelectric applications to increase the thermoelectric figure of merit, $zT$.

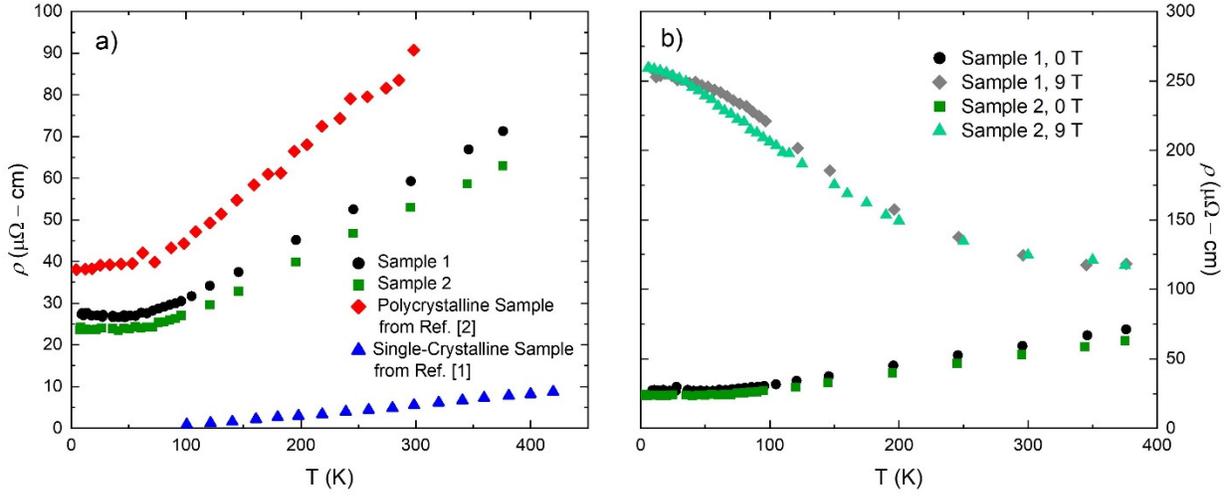

*Figure 4. (a) Temperature-dependence of electrical resistivity at 0 T for Samples 1 and 2 compared to previously published data. (b) Temperature-dependence of electrical resistivity for Samples 1 and 2 at 0 T and 9 T.*

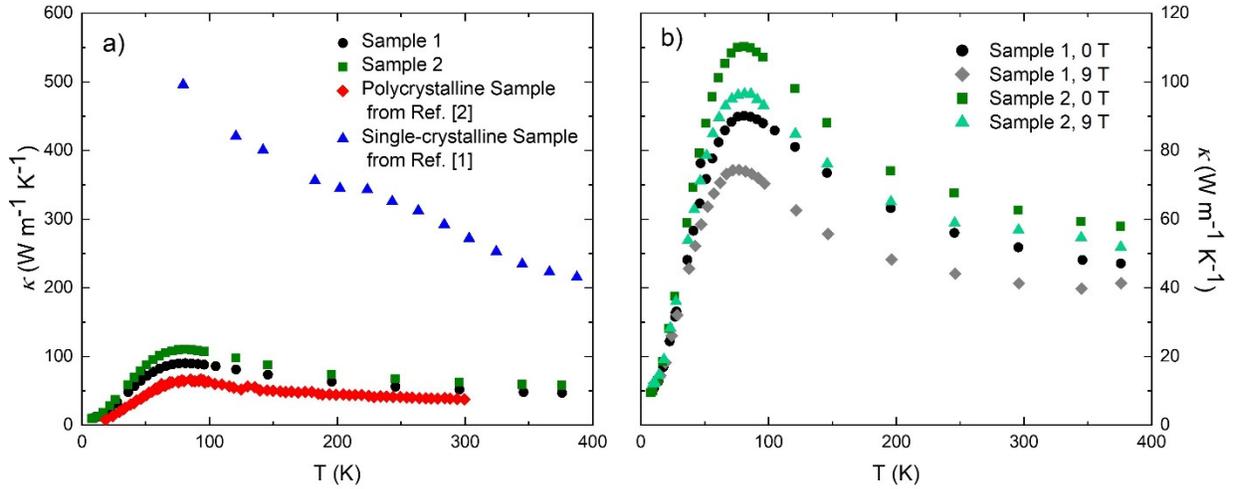

*Figure 5. (a) Temperature-dependence of thermal conductivity at 0 T for Samples 1 and 2 compared to previously published data. (b) Temperature-dependence of thermal conductivity for Samples 1 and 2 at 0 T and 9 T.*

## IV. Theoretical Results and Discussion

A combined look at the key transport values for Samples 1 and 2, compared to previously published data for NbP, is shown in Table I.

*Table I. Comparison of maximum values of key TE transport phenomena and corresponding temperatures at which they occur*

| NbP Sample | Maximum Magnitude of $S_{xx}$ at 0 T | | Maximum Magnitude of $S_{xxz}$ at 9 T | | Maximum $S_{xyz}$ at 9 T | |
|---|---|---|---|---|---|---|
| | $S_{xx}$ | $T$ | $S_{xxz}$ | $T$ | $S_{xyz}$ | $T$ |
| Sample 1 | 43.3 µV K$^{-1}$ | 64.50 K | 103.8 µV K$^{-1}$ | 105.4 K | 51.6 µV K$^{-1}$ | 159.1 K |
| Sample 2 | 26.7 µV K$^{-1}$ | 102.5 K | 59.1 µV K$^{-1}$ | 94.77 K | 99.7 µV K$^{-1}$ | 163.8 K |
| Polycrystalline NbP from Ref. [2] | 20 µV K$^{-1}$ | 95 K | 50 µV K$^{-1}$ | 110 K | 92 µV K$^{-1}$ | 136 K |
| Single-Crystalline NbP from Ref. [1] | 8 µV K$^{-1}$ | 100 K | -- | -- | 800 µV K$^{-1}$ | 109 K |

Ref. [1] determined that as the chemical potential moved with temperature, it eventually approached the energy of the Weyl points, where it became pinned due to a minimum density of states in the Dirac bands. In a Type I WSM, the energy of the Weyl points is also the location where NbP is fully compensated. In single-crystalline NbP, both the thermopowers $S_{xx}$ and $S_{xyz}$ were maximized (in magnitude) at a temperature T ≈ 100 K, which is close to the temperature at which the chemical potential begins to shift strongly toward the energy of the Weyl points. In the polycrystalline Samples 1 and 2, the maximum magnitude of $S_{xx}$ for Sample 1 occurs at 64.50 K and the maximum magnitude for Sample 2 occurs at 102.5 K, differing from the single crystal. The maximum $S_{xyz}$ from the Nernst effect, though, occurs at larger temperatures, with the maximum $S_{xyz}$ of Sample 1 occurring at 159.1 K and the maximum of Sample 2 occurring at 163.8 K. The temperatures at which these maxima occur differ between samples, indicating that the samples encompass differing temperature dependences of the chemical potential and thus have different Fermi levels. Because the Fermi level can be manipulated via doping, this difference offers evidence that both the maximum magnitude of thermopower and the temperature at which it occurs can be tuned via doping.

In order to provide a theoretical interpretation of our results, we perform calculations of both $S_{xxz}$ and $S_{xyz}$ arising from a Weyl electron band. These calculations assume an ideal, gapless, linear Weyl dispersion with constant Fermi velocity $v_F$. Our calculations use the Boltzmann transport description, which is valid in the semiclassical limit where the electron system is far from the extreme quantum limit of magnetic field. A detailed description of these calculations is provided in the Supplementary Materials [22]. The calculations take as an input the value of the (zero-temperature) Fermi Energy $E_F$ (which in general is modified by doping), the transport scattering time $\tau$, and the Fermi velocity $v_F$. Except where otherwise noted, the theory curves presented in this section use the values of $E_F$ and $\tau$ that were estimated for the single-crystalline samples in Ref. [1] [$E_F \approx$ 8 meV = $k_B \times$ (92.8 K) and $\tau = $ 0.3 ps, respectively], and we have not attempted to fit these parameters to the data. For simplicity, we use a description where $\tau$ is an energy-independent constant and we take $v_F$ to be isotropic and equal to $10^5$ m s$^{-1}$. In the Supplementary Materials [22], we comment on the effect of energy dependence of the scattering rate; in general, the scattering rate alters certain numeric prefactors but does not change any results qualitatively.

At zero magnetic field, and within our approximation of a constant scattering time, the thermopower is given by equation (1):

$$S_{xx} = -\frac{k_B}{e} \frac{2\pi^2 k_B T \mu}{3\mu^2 + \pi^2 (k_B T)^2} \tag{1}$$

where $\mu = \mu(T)$ is the temperature-dependent chemical potential (a full expression for $\mu(T)$ for an isolated Weyl band is given in the Supplementary Materials [22]) with 0 eV defined as the energy of the Weyl points, $k_B$ is the Boltzmann constant, and $e$ is the elementary charge. At low temperatures, the chemical potential approaches a constant value equal to the Fermi energy $E_F = \hbar v_F (6\pi^2 n/g)^{1/3}$, where $\hbar$ is the reduced Planck constant, $n$ is charge carrier density at $T = 0$, and $g$ is the degeneracy. Consequently, at low temperatures the magnitude of the thermopower increases linearly with temperature, $S_{xx} \simeq -(k_B/e)(2\pi^2/3)(T/T_F)$, where $T_F \equiv E_F/k_B$ is the Fermi temperature. Such growth can be understood, as in usual metals, as a consequence of the linear growth in entropy of the electron system as the temperature is increased due to thermal excitations of electrons around the Fermi surface. On the other hand, at high temperatures $T \gg T_F$, the chemical potential for an isolated Weyl band falls toward zero as $\mu \propto 1/T^2$, and consequently the magnitude of the thermopower decreases as $1/T^3$. This decline at large temperatures is a result of thermally-excited holes appearing in the valence band, which cancel the thermopower contribution from electrons. Equation (1) predicts a universal dependence of the thermopower on the ratio $T/T_F$, with $S_{xx}$ having a maximum magnitude of $\approx$ 156 $\mu$V K$^{-1}$ at a temperature $T \approx 0.35 T_F$. Thus, the prediction of this calculation is that by changing the doping (and therefore the value of $T_F$), one can set the temperature corresponding to the thermopower maximum, while the value of the maximum peak thermopower is unaffected. This dependence is shown in Figure 6. The experimental data in Figure 1 shows a somewhat smaller value of the maximum thermopower, which varies from one sample to the next. This discrepancy may arise from the presence of nearby trivial bands, as we discuss below.

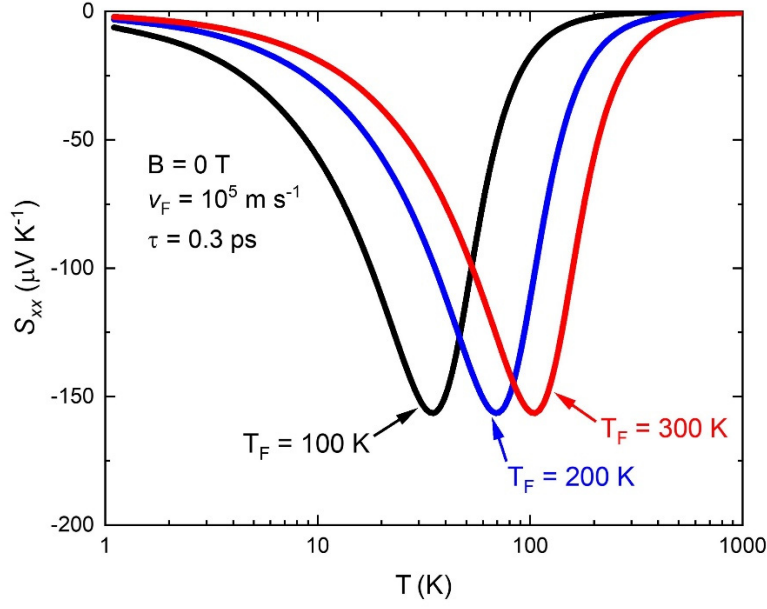

*Figure 6. Theoretical model of $S_{xx}$ at various $T_F$ and 0 T, where $v_F = 10^5$ m s$^{-1}$ and t = 0.3 ps.*

The experimental data in Figure 1 is consistent with the theoretical result of Equation (1) at low temperature. Increased doping reduces the slope of $S_{xx}$ versus $T$. At temperatures higher than $\approx 100$ K, however, $S_{xx}$ declines in magnitude rapidly toward zero even for samples with large $T_F$, and the magnitude of $S_{xx}$ never reaches values as high as the maximum magnitude of $\approx 156\ \mu$V K$^{-1}$ as predicted above. At room temperature or higher the sign of the thermopower is inverted. One possible origin for this discrepancy with the theory relates to the presence of topologically trivial bands, in addition to the Weyl bands dominating transport, in the band structure, which have been predicted based on first-principles calculations [23]. For example, if a trivial *n*-type band is nearby in energy to the Weyl points, then at elevated temperatures this trivial band has an increased concentration of thermally-populated electrons, such that the chemical potential is forced to move below the Weyl point in order to maintain charge neutrality. In this scenario, when the chemical potential crosses through the energy of the Weyl points, the thermopower arising from the Weyl bands rapidly falls to zero, and at higher temperatures the sign of the thermopower is inverted (along with the sign of the Hall coefficient, as we discuss in the Supplementary Materials [22]).

An applied magnetic field tends to enhance the thermopower via the magneto-Seebeck effect. A theoretical calculation for an isolated Weyl band in the limit $T \ll T_F$ gives Equation (2):

$$S_{xxz} \simeq -\frac{k_B}{e}\frac{\pi^2}{3}\frac{2 + 3(B/B_0)^2}{1 + (B/B_0)^2}\frac{T}{T_F} \qquad (2)$$

where $B_0 = E_F/ev_F^2\tau$. (For reference, at $T_F \sim 100$ K, $v_F \sim 10^5$ m s$^{-1}$, and $\tau \sim 1$ ps, the magnetic field scale $B_0 \sim 10$ T.) At sufficiently large magnetic fields $B \gg B_0$, the slope of $S_{xxz}$ versus temperature is increased by a factor of 3/2. At temperatures comparable to $T_F$ the enhancement by magnetic field is more dramatic. This enhanced magneto-thermopower is illustrated in Figure 7.

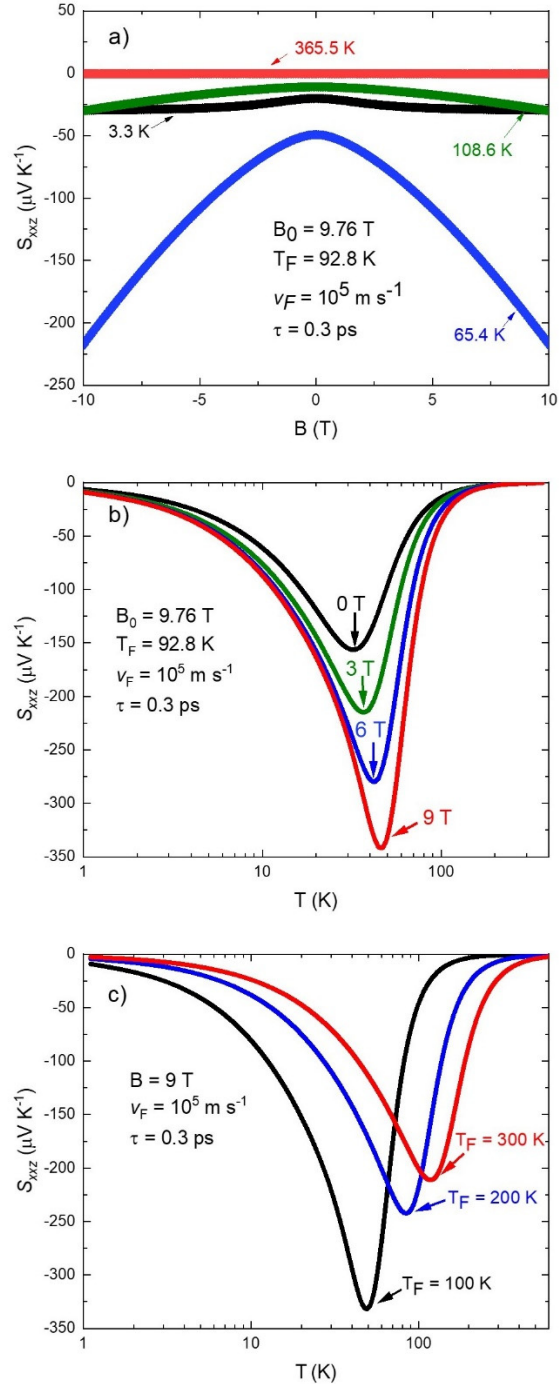

Figure 7. Theoretical calculations of: (a) $S_{xxz}$ as a function of magnetic field at various temperatures, with constant $T_F = 92.8$ K and $B_0 = 9.76$ T; (b) $S_{xxz}$ as a function of temperature at discrete magnetic fields,

with constant $T_F$ = 92.8 K, and $B_0$ = 9.76 T; and (c) $S_{xxz}$ at various $T_F$ and constant B = 9 T. For all plots, $v_F = 10^5$ m s$^{-1}$ and t = 0.3 ps.

When the temperature becomes comparable to $T_F$, the chemical potential drops toward the energy of the Weyl point, and $S_{xxz}$ falls again as in the zero-field case of $S_{xx}$. In principle, the growth of $S_{xxz}$ with temperature can continue toward larger temperatures $T \gtrsim T_F$ in cases where the magnetic field is strong enough to produce a large Hall angle ($r_{xyz} > r_{xxz}$) [18, 19, 24]. In this regime, the peak magneto-thermopower is parametrically enhanced by the magnetic field, but this large magnetic field regime is not realized in our experiments

Finally, we can use our theory to calculate the dependence of the Nernst thermopower, $S_{xyz}$, on temperature and doping. In general, the magnitude, and even the sign, of the Nernst thermopower depends sensitively on the scattering mechanism and its dependence on the quasiparticle energy. Since we are using a simple model with energy-independent scattering, our theory should be taken only as a general indication of the dependence on temperature and doping. Within this model $S_{xyz}$ to lowest order in magnetic field can be calculated as:

$$S_{xyz} \simeq \frac{k_B}{e} \frac{\pi^2}{3} \frac{B}{B_0} \frac{E_F^2 \left(\frac{\pi^2}{3} k_B^2 T^2 - \mu^2\right)}{\left(\frac{\pi^2}{3} k_B^2 T^2 + \mu^2\right)^2} \frac{T}{T_F} \qquad (3)$$

(Notice that $S_{xyz}$ remains finite in the limit of zero chemical potential and finite temperature, where both $E_F$ and $\mu$ go to zero, since the values of $B_0$ and $T_F$ are proportional to $E_F$.) As the temperature is increased, the value of $S_{xyz}$ first increases due to the increased entropy of the charge carriers. At temperatures larger than ~ $T_F$, however, $S_{xyz}$ falls again as $1/T$. Ignoring the presence of any possible trivial bands and inserting the corresponding behavior of $\mu(T)$ gives $S_{xyz}$ which at low field has a maximum at $T \approx 0.6 T_F$. At large magnetic fields $B \gg B_0$, $S_{xyz}$ falls as $1/B$. This behavior of $S_{xyz}$ is shown in Figure 8.

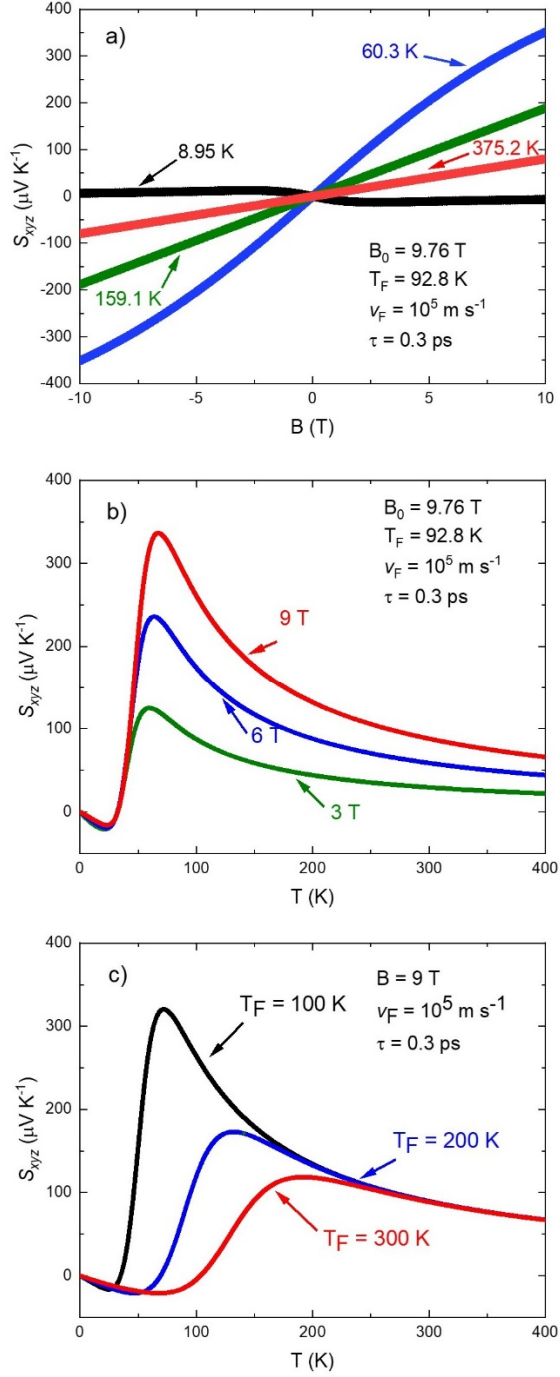

*Figure 8. Theoretical calculations of: (a) $S_{xyz}$ as a function of magnetic field at various temperatures, with constant $T_F = 92.8$ K, and $B_0 = 9.76$ T; (b) $S_{xyz}$ as a function of temperature at discrete magnetic fields, with constant $T_F = 92.8$ K, and $B_0 = 9.76$ T; and (c) $S_{xyz}$ at various $T_F$ and constant $B = 9$ T. For all plots, $v_F = 10^5$ m s$^{-1}$ and $t = 0.3$ ps.*

## V. Future Outlook

The simultaneous presence of both a large Nernst effect and a magneto-Seebeck effect, where the thermopowers from each effect are on the same order of magnitude, is not typically found in a single material. In a two-carrier system, the two signs of charge carrier typically counteract each other in a longitudinal geometry as both polarities of charge carriers condense on the cold side. However, the unique band structure of WSMs leads to a large longitudinal TE effect in the presence of a transverse magnetic field [18]. This result implies that, in the presence of a magnetic field, the total effective thermopower of NbP can include contributions from both longitudinal and transverse TE effects, which effectively doubles the total thermopower if a TE device is setup appropriately. Thermoelectric devices using WSMS typically utilize the transverse geometry exclusively due to their large Nernst effect. Yet in WSMs, a large magneto-Seebeck effect also exists concurrently but is unused in transverse devices. Because both transverse and longitudinal effects are simultaneously present in NbP, a thermoelectric device utilizing both could be not only more efficient but also reasonably simplistic. A single leg schematic of such a proposed device is shown in Figure 9, demonstrating how both the longitudinal and transverse voltage output could be harnessed simultaneously. However, geometrical design optimization would still be necessary for a heterojunction device, where both n-type and p-type TE legs are necessary to return the longitudinal electric field to an isothermal plane for a total longitudinal output voltage. The device would need to be wired in such a way that the voltage output of both magneto-Seebeck and Nernst voltages are additive, with only one magnetic field direction being needed for the entire device. We offer Figure 9 as a concept only, as such a device would need further optimization to operate appropriately.

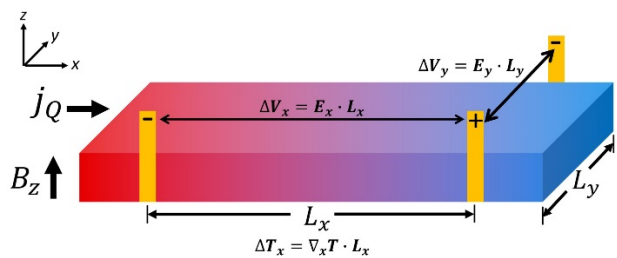

*Figure 9. Schematic of a single TE leg for the proposed device utilizing contributions from both the Nernst and Seebeck effects.*

By calculating the thermoelectric figure of merit, $zT$, and power factor, $PF$, the performance enhancement of such a device can be quantified. Calculating $zT$ differs depending on the geometry of the TE transport property used. Because both the magneto-Seebeck and Nernst effects are being utilized, two different $zT$s and $PF$s can be calculated. Equations (4) and (5) are the formulae used for calculating longitudinal $zT$ and $PF$, respectively, from the longitudinal thermopower $S_{xxz}$:

$$zT_{xxz} = \frac{S_{xxz}^2 \sigma_{xxz}}{\kappa_{xxz}} T \quad (4)$$

$$PF_{xxz} = S_{xxz}^2 \sigma_{xxz} \quad (5)$$

Similarly, the method for calculating transverse *zT* and *PF* is shown in Equations (6) and (7), respectively, using the transverse thermopower $S_{xyz}$:

$$zT_{xyz} = \frac{S_{xyz}^2 \sigma_{yyz}}{\kappa_{xxz}} T \quad (6)$$

$$PF_{xyz} = S_{xyz}^2 \sigma_{yyz} \quad (7)$$

It should be noted that electrical resistivity $\rho$, which was directly measured, is not mathematically the direct inverse of electrical conductivity $\sigma$ but a tensor making $\sigma_{xxz} = \frac{\rho_{xxz}}{\rho_{xxz}^2 + \rho_{xyz}^2}$. However, because the Hall resistivity $\rho_{xyz}$ is significantly smaller than the longitudinal electrical resistivity $\rho_{xxz}$ [22], $\sigma_{xxz} \approx \frac{1}{\rho_{xxz}}$. Also, because the samples are polycrystalline, it can reasonably be assumed that the electrical resistivity is isotropic such that $\rho_{xxz} \approx \rho_{yyz}$.

We now define the effective *zT* and *PF* produced by the combined use of Nernst and magneto-Seebeck effects using Equations (8) and (9), respectively:

$$zT_{effective} = \frac{(|S_{xyz}| + |S_{xxz}|)^2 \sigma_{xxz}}{\kappa_{xxz}} T \quad (8)$$

$$PF_{effective} = (|S_{xyz}| + |S_{xxz}|)^2 \sigma_{xxz} \quad (9)$$

Here, we add the magnitude of the thermopowers assuming that such a device would be wired in a way such that the voltages are all the same sign when combined. Equations (8) and (9) refer to the $zT_{effective}$ and $PF_{effective}$ for a single leg of the proposed device, as shown in Figure 9. However, because a conventional TE device requires two legs (n-type and p-type pair), the actual effective thermopower of the device would be the sum of $|S_{xxz}|$ and $|S_{xyz}|$ from each of the legs. Thus, both a transverse and longitudinal voltage would come from each of the n-type and p-type legs. Furthermore, geometrical device optimization, specifically in the leg cross-sectional area, is not considered here. An advantage of transverse TE devices is that the output scales extrinsically with the device size (i.e. a larger voltage output would result from increasing the device size in the parallel dimension), but longitudinal devices require an optimization in leg cross-sectional area to optimize *zT*. Since both longitudinal and transverse TE effects are explored in this device, geometrical optimization would be necessary to realistically design such a device.

The resulting *zT* and *PF* values of Samples 1 and 2 from Equations (4) through (9) are shown in Figure 10.

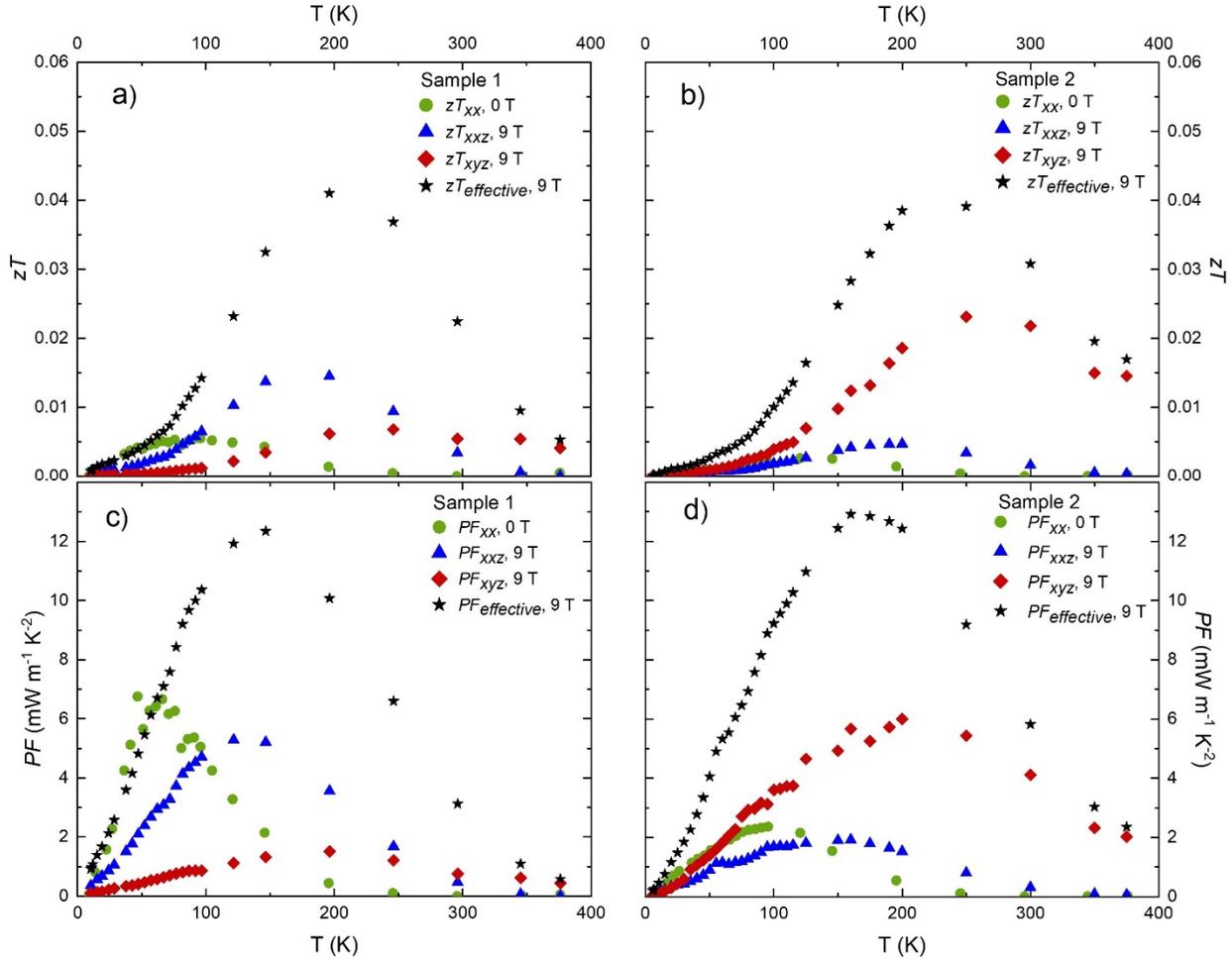

*Figure 10. Temperature-dependence of zT and PF in Sample 1 (frames (a) and (b), respectively) and Sample 2 (frames (c) and (d) respectively) with contributions from the Seebeck effect, magneto-Seebeck effect at 9 T, Nernst effect at 9 T, and combined longitudinal and transverse effects at 9 T for the proposed device.*

$zT_{xxz}$ and $PF_{xxz}$ refer to the longitudinal figure of merit and power factor calculated from the Seebeck effect (Equations (4) and (5)), and $zT_{xyz}$ and $PF_{xyz}$ refer to the transverse figure of merit and power factor calculated from the Nernst effect (Equations (6) and (7)). $zT_{effective}$ and $PF_{effective}$ were calculated using the isothermal data for $S_{xxz}$ and $S_{xyz}$, shown in Figures 2 and 3 respectively, both of which were measured independently and isothermally. The proposed device design in Figure 9 is inherently adiabatic, therefore the data shown in Figure 10 should be considered as an estimate only.

When considering all three independent TE effects (Seebeck, magneto-Seebeck, and Nernst effects) observed in this work for Sample 1, $zT_{xxz}$ (9 T) has the largest magnitude of 0.0145 at 196.4 K, which reflects the large magneto-Seebeck effect observed in this material. $PF_{xx}$ (0 T) reached the largest magnitude of 6.7 mW m$^{-1}$ K$^{-2}$ at 47 K. Unfortunately, this power factor drops off to near zero above 200 K. However, $PF_{xxz}$ (9 T) reaches a lower maximum of 5.25 mW m$^{-1}$ K$^{-2}$ at 121 K but does not drop to near zero until ~300 K.

Of the TE effects observed in Sample 2, $zT_{xyz}$ (9 T) has the largest magnitude of 0.023 at 250.0 K, which reflects the larger Nernst effect observed in this sample. $PF_{xyz}$ (9 T) reached the largest magnitude of 5.998 mW m$^{-1}$ K$^{-2}$ at 200.0 K. A large $PF_{xyz}$ (9 T) is maintained from ~100 K to ~300 K.

When utilizing both the magneto-Seebeck and Nernst effects simultaneously, the effective thermopower is increased in the proposed device while maintaining the same electrical resistivity and thermal conductivity. This results in $zT_{effective}$ (9 T) reaching a maximum of 0.041 at 196.4 K, which is nearly triple in value as compared to the highest figure of merit for an independent TE effect in Sample 1. Similarly, $PF_{effective}$ (9 T) reaches a maximum of 12.35 mW m$^{-1}$ K$^{-2}$ at 146.5 K, nearly double the maximum power factor for an independent TE effect in Sample 1, and the large power factor is maintained across a large temperature range of approximately 100 K to 200 K.

The $zT_{effective}$(9 T) of Sample 2 reaches a maximum of 0.039 at 250.0 K, which is not significantly different to the Sample 1 maximum value, however it occurs at a much higher temperature. Sample 2's $PF_{effective}$ (9 T) reaches a maximum of 12.91 mW m$^{-1}$ K$^{-2}$ at 160.0 K. Once again, the magnitude does not vary significantly from that of Sample 1, but the temperature at which it occurs has changed. This indicates that tuning the operating temperature of the device by means of tuning $T_F$ via doping will not significantly affect the magnitude of $zT$ or $PF$ of the device.

The $zT_{effective}$ (9 T) found here for the proposed device is competitive with, and even exceeds, figures of merit found in recent works on TE transport in WSMs. Table II shows the largest $zT$ reported experimentally for other WSMs, where the maximum $zT$ observed by a single TE effect (either Seebeck, magneto-Seebeck, or Nernst effect independently) is $zT_{xx}$(0 T) = 0.042 at ~80 K in Mo$_{0.92}$Nb$_{0.08}$Te$_2$ [25], which is comparable to the $zT_{effective}$ of the NbP samples in this work. Similar to the results for polycrystalline NbP reported in this work, MoTe$_2$ demonstrates significant changes in the maximum $zT$ and temperature at which that maximum occurs – $zT_{xx}$(0 T) = 0.00297 at ~20 K for MoTe$_2$ and $zT_{xx}$(0 T) = 0.042 at 80 K for Mo$_{0.92}$Nb$_{0.08}$Te$_2$ [25] – offering further evidence that doping can be used as a tuning mechanism to increase $zT$.

*Table II. Comparison of zT values from independent TE effects in WSMs*

| Material | Sample Type | Max zT | Conditions | Reference |
|---|---|---|---|---|
| NbP, Sample 1 | Polycrystalline | 0.0055 | $zT_{xx}$, (0 T), 95.78 K | this work |
| NbP, Sample 1 | Polycrystalline | 0.0145 | $zT_{xxz}$, (9 T), 196.4 K | this work |
| NbP, Sample 1 | Polycrystalline | 0.0068 | $zT_{xyz}$, (9 T), 246.2 K | this work |
| **NbP, Sample 1** | **Polycrystalline** | **0.041** | $zT_{effective}$, **(9 T), 196.4 K** | **this work** |
| NbP, Sample 2 | Polycrystalline | 0.0027 | $zT_{xx}$, (0 T), 125.0 K | this work |
| NbP, Sample 2 | Polycrystalline | 0.0047 | $zT_{xxz}$, (9 T), 200.0 K | this work |
| NbP, Sample 2 | Polycrystalline | 0.023 | $zT_{xyz}$, (9 T), 250.0 K | this work |
| **NbP, Sample 2** | **Polycrystalline** | **0.039** | $zT_{effective}$, **(9 T), 250.0 K** | **this work** |
| NbP | Polycrystalline | 0.0098 | $zT_{xyz}$, (9 T), 140 K | [2] |
| $Co_3Sn_2S_2$ | Single Crystal | 0.0221 | $zT_{xx}$ (0 T), 180 K | [3] |
| TaAs | Single Crystal | 0.0135 | $zT_{xx}$ (0 T), 220 K | [26] |
| TaP | Single Crystal | 0.003252 | $zT_{xx}$ (0 T), 75 K | [27] |
| $MoTe_2$ | Single Crystal | 0.00297 | $zT_{xx}$ (0 T), 20 K | [25] |
| $Mo_{0.92}Nb_{0.08}Te_2$ | Single Crystal | 0.042 | $zT_{xx}$ (0 T), 80 K | [25] |
| $YbMnBi_2$ | Single Crystal | 0.003252 | $zT_{xyz}$ (2 T), 180 K | [9] |

As such, the proposed device offers a pathway to make polycrystalline WSMs competitive for magnetic TE device applications. Worthy of note is that the majority of TE transport literature focuses on single-crystalline samples. Because this work shows a significant difference between the transport properties of polycrystalline NbP compared to single-crystalline NbP, most notably a large isothermal magneto-Seebeck effect present in polycrystalline but not single-crystalline samples, there may be potential for TE applications in the other WSMs shown in Table II in their polycrystalline form.

Both the experimental and theoretical findings presented in the previous sections indicate that the TE properties of NbP are highly sensitive to changes in $T_F$ and thus $E_F$, as TE transport is strongly dependent on the position of the chemical potential relative to the energy of the Weyl points. This sensitivity implies the possibility of tuning $E_F$ of NbP and other Type I WSMs to optimize TE properties at desirable operating conditions for devices. For the new proposed device of Figure 9, maximizing the sum of $|S_{xxz}|$ and $|S_{xyz}|$ at the same temperature and magnetic field is desirable for ultimately maximizing $zT_{effective}$. This can be achieved by tuning $T_F$ in a material to where the maximum sum of $|S_{xxz}|$ and $|S_{xyz}|$ is observed based on desired device operating conditions of temperature and magnetic field. While the experimental data shows only a comparison of two different Fermi temperatures, a theoretical model of the change in combined thermopower through deliberate tuning is shown in Figure 11.

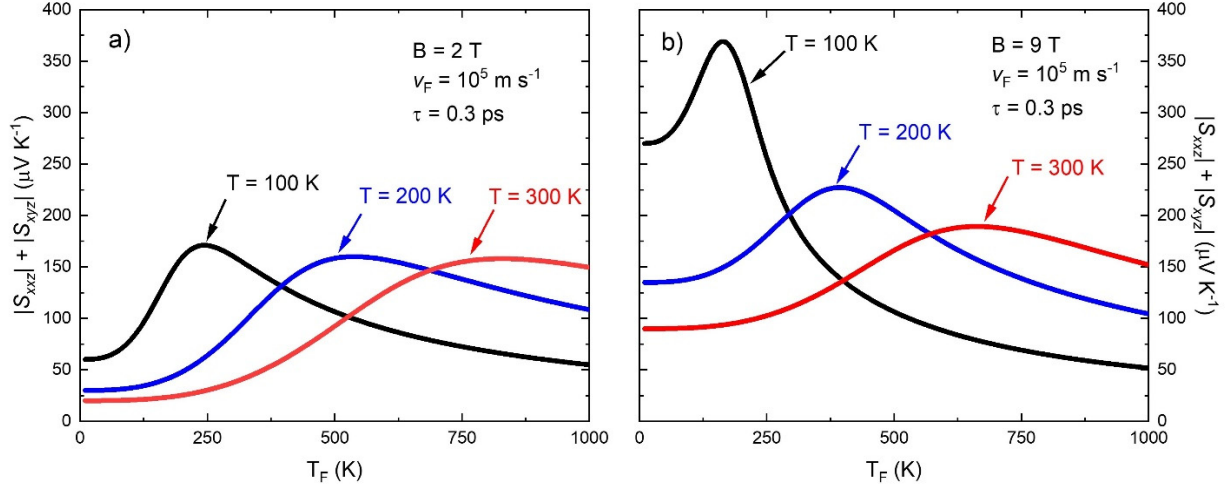

*Figure 11. Theoretical calculation of combined magnetic thermopowers ($|S_{xxz}| + |S_{xyz}|$) as a function of $T_F$ at two externally applied discrete fields: (a) 2 T and (b) 9 T.*

Figure 11 indicates that the sum $|S_{xxz}| + |S_{xyz}|$ depends significantly on $T_F$, the externally applied magnetic field, and the temperature. Again, we offer doping as a tuning mechanism to maximize the TE transport of a magnetic WSM TE device, where an appropriate value of $T_F$ can be chosen to fit an application's operating magnetic field and temperature constraints.

## VI. Conclusions

The TE properties of WSMs have been investigated recently as they offer insight into fundamental topological transport, but WSMs also innately demonstrate promise as TE materials themselves. Single crystals are typically considered to have the most impressive TE transport properties; however, single crystals are also costly, difficult to synthesize, and typically small and fragile. If WSMs are to be used in practical applications, polycrystalline samples are the cheaper and more durable option. Here, we find a simultaneously large magneto-Seebeck and Nernst effect in two polycrystalline samples of NbP, with a significantly reduced thermal conductivity from that of single-crystalline values and reasonably low electrical resistivity, further indicating that magneto-thermoelectric effects are an area of major interest in WSMs. When comparing experimental data to theoretical models, we have determined that TE effects in NbP are highly sensitive to doping, altering not only the maximum values of TE transport properties but the temperatures at which these maxima occur. We conclude that changes in the Fermi temperature through doping can be used as a sensitive tuning mechanism for TE properties and the temperatures at which they are maximized. Finally, because NbP possesses the unique simultaneous presence of both a large magneto-Seebeck and Nernst effect, we propose a device which utilizes both thermopowers produced in the presence of a magnetic field. The resulting $zT$ of such a device was found to be significantly higher than the $zT$ observed by the individual TE effects and is greater than *zT* found in other competitive WSMs. We offer all of this as evidence for the viability of polycrystalline WSMs in TE devices.

**Acknowledgements**

EFS and SJW acknowledge support from the U.S. Department of Energy, Office of Science, Office of Basic Energy Sciences Early Career Research Program under Award Number DE-SC0020154. BS was supported in part by the NSF under Grant No. DMR-2045742. C. Fu acknowledges support from the Fundamental Research Funds for the Central Universities.

# References

[1] S. J. Watzman, T. M. McCormick, C. Shekhar, S. C. Wu, Y. Sun, A. Prakash, C. Felser, N. Trivedi, and J. P. Heremans. Dirac dispersion generates unusually large Nernst effect in Weyl semimetals. *Phys. Rev. B* **97**, 161404(R) (2018).

[2] C. Fu, S. N. Guin, S. J. Watzman, G. Li, E. Liu, N. Kumar, V. Su, W. Schnelle, G. Auffermann, C. Shekhar, Y. Sun, J. Gooth, and C. Felser. Large Nernst power factor over a broad temperature range in polycrystalline Weyl semimetal NbP. *Energy Environ. Sci*. **11**, 2813 (2018).

[3] S. N. Guin, P. V., Y. Zhang, N. Kumar, S. J. Watzman, C. Fu, E. Liu, K. Manna, W. Schnelle, J. Gooth, C. Shekhar, Y. Sun, and C. Felser. Zero-Field Nernst Effect in a Ferromagnetic Kagome-Lattice Weyl-Semimetal $Co_3Sn_2S_2$. *Adv. Mater*. **31**, 1806622 (2019).

[4] R. Modak, K. Goto, S. Ueda, Y. Miura, K. Uchida, and Y. Sakuraba. Combinatorial tuning of electronic structure and thermoelectric properties in $Co_2MnAl_{1-x}Si_x$ Weyl semimetals. *APL Mater*. **9**, 031105 (2021).

[5] B. Peng, H. Zhang, H. Shao, H. Lu, D. W. Zhang, and H. Zhu. High thermoelectric performance of Weyl semimetal TaAs. *Nano Energy* **30** (2016).

[6] S. N. Guin, K. Manna, J. Noky, S. J. Watzman, C. Fu, N. Kumar, W. Schnelle, C. Shekhar, Y. Sun, J. Gooth, and C. Felser. Anomalous Nernst effect beyond the magnetization scaling relation in the ferromagnetic Heusler compound $Co_2MnGa$. *NPG Asia Materials* **11** (2019).

[7] M. Ikhlas, T. Tomita, T. Koretsune, M. Suzuki, D. Nishio-Hamane, R. Arita, Y. Otani, and S. Nakatsuji. Large anomalous Nernst effect at room temperature in a chiral antiferromagnet. *Nat. Phys*. **13** (2017).

[8] X. Li, L. Xu, L. Ding, J. Wang, M. Shen, X. Lu, Z. Zhu, and K. Behnia. Anomalous Nernst and Righi-Leduc Effects in $Mn_3Sn$: Berry Curvature and Entropy Flow. *Phys. Rev. Lett*. **119**, 056601 (2017).

[9] Y. Pan, C. Le, B. He, S. J. Watzman, M. Yao, J. Gooth, J. P. Heremans, Y. Sun, and C. Felser. Giant anomalous Nernst signal in the antiferromagnet $YbMnBi_2$. *Nat. Mater*. **21,** 203–209 (2022).

[10] H. Wang, X. Luo, W. Chen, N. Wang, B. Lei, F. Meng, C. Shang, L. Ma, T. Wu, X Dai, Z. Wang, and X. Chen. Magnetic-field enhanced high-thermoelectric performance in topological Dirac semimetal $Cd_3As_2$ crystal. *Sci. Bull*. **63**, 411 (2018).

[11] J. Xiang, S. Hu, M. Lyu, W. Zhu, C. Ma, Z. Chen, F. Steglich, G. Chen, and P. Sun. Large transverse thermoelectric figure of merit in a topological Dirac semimetal. *Sci. China-Phys. Mech. Astron*. **63**, 237011 (2020).

[12] J. Hu, M. Caputo, E. Bonini Guedes, S. Tu, E. Martino, A. Magrez, H. Berger, J. H. Dil, H. Yu, and J. Anserme. Large magnetothermopower and anomalous Nernst effect in $HfTe_5$. *Phys. Rev. B* **100**, 115201 (2019).

[13] K. Tsuruda, K. Nakagawa, M. Ochi, K. Kuroki, M. Tokunaga, H. Murakawa, N. Hanasaki, and H. Sakai. Enhancing Thermopower and Nernst Signal of High-Mobility Dirac Carriers by Fermi Level Tuning in the Layered Magnet $EuMnBi_2$. *Adv. Funct. Mater*., **31**, 2102275 (2021).

[14] R. Lundgren, P. Laurell, and G. A. Fiete. Thermoelectric properties of Weyl and Dirac semimetals. *Phys. Rev. B* **90**, 165115 (2014).

[15] G. Sharma, P. Goswami, and S. Tewari. Nernst and magnetothermal conductivity in a lattice model of Weyl fermions. *Phys. Rev. B* **93** 035116 (2016).

[16] E. H. Putley. *The Hall Effect and Related Phenomena.* Butterworths, London, UK (1960).

[17] S. Boona, H. Jin, and S. J. Watzman. Transverse thermal energy conversion using spin and topological structures. *J. Appl. Phys*. **130**, 171101 (2021).

[18] B. Skinner and L. Fu. Large, nonsaturating thermopower in a quantizing magnetic field. *Sci. Adv*. **4** (5) (2018).

[19] X. Feng and B. Skinner. Large enhancement of thermopower at low magnetic field in compensated semimetals. *Phys. Rev. Materials* **5**, 024202 (2021).

[20] W. Liu, Z. Wang, J. Wang, H. Bai, Z. Li, J. Sun, X. Zhou, J. Luo, W. Wang, C. Zhang. J. Wu. Y. Sun, Z. Zhu, Q. Zhang, and X. Tang. Weyl Semimetal States Generated Extraordinary Quasi-Linear Magnetoresistance and Nernst Thermoelectric Power Factor in Polycrystalline NbP. *Adv. Funct. Mater*. 2202143 (2022).

[21] C. Shekhar, A. K. Nayak, Y. Sun, M. Schmidt, M. Nicklas, I. Leermakers, U. Zeitler, Y. Skourski, J. Wosnitza, Z. Liu, Y. Chen, W. Schnelle, H. Borrmann, Y. Grin, C. Felser, and B. Yan. Extremely large magnetoresistance and ultrahigh mobility in the topological Weyl semimetal candidate NbP. *Nat. Phys*. **11** (2015).

[22] See Supplementary Materials for this manuscript.

[23] H. Weng, C. Fang, Z. Fang, B. A. Bernevig, and X. Dai. Weyl Semimetal Phase in Noncentrosymmetric Transition-Metal Monophosphides. *Phys. Rev. X* **5**, 011029 (2015).

[24] V. Kozii, B. Skinner, and L. Fu. Thermoelectric Hall conductivity and figure of merit in Dirac/Weyl materials. *Phys. Rev. B* **99**, 155123 (2019).

[25] H. Sakai, K. Ikeura, M. S. Bahramy, N. Ogawa, D. Hashizume, J. Fujioka, Y. Tokura, and S. Ishiwata. Critical enhancement of thermopower in a chemically tuned polar semimetal $MoTe_2$. *Sci. Adv*. **2**, e1601378 (2016).

[26] J. Xiang, S. Hu, M. Lv, J. Zhang, H. Zhao, G. Chen, W. Li, Z. Chen, and P. Sun. Anisotropic thermal and electrical transport of Weyl semimetal TaAs. *J. Phys.: Condens. Matter* **29**, 485501 (2017).

[27] F. Han, N Andrejevic, T. Nguyen, V. Kozii, Q. T. Nguyen, T. Hogan, Z. Ding, R. Pablo-Pedro, S. Parjan, B. Skinner, A. Alatas, E. Alp, S. Chi, J. Fernandez-Baca, S. Huang, L. Fu, and M. Li. Quantized thermoelectric Hall effect induces giant power factor in a topological semimetal. *Nat. Commun*. **11**, 6167 (2020).

**Supplementary Materials: Doping as a tuning mechanism for magneto-thermoelectric effects to improve *zT* in polycrystalline NbP**

**Authors: Eleanor F. Scott[1], Katherine A. Schlaak[2], Poulomi Chakraborty[3], Chenguang Fu[4,5], Satya N. Guin[5,6], Safa Khodabakhsh[1], Ashley E. Paz y Puente[1], Claudia Felser[5], Brian Skinner[3], Sarah J. Watzman[1*]**

1. Department of Mechanical and Materials Engineering, University of Cincinnati, Cincinnati, OH 45221
2. Department of Physics, University of Cincinnati, Cincinnati, OH 45221
3. Department of Physics, The Ohio State University, Columbus, OH 43210
4. Department of Materials Science and Engineering, Zhejiang University, Hangzhou, China 310027
5. Max Planck Institute for Chemical Physics of Solids, Dresden, Germany 01187
6. Department of Chemistry, Birla Institute of Technology and Science, Pilani – Hyderabad Campus, Hyderabad, India 500078
*Please direct correspondence to watzmasj@ucmail.uc.edu

**I. Sample Characterization**

Sample 1 was cut from the ingot and left unannealed, synthesized as described in the main text. Sample 2 was wrapped in Nb foil with excess NbP powder then encapsulated in a quartz tube in an argon environment for annealing. The NbP powder was added to prevent loss of P from the sample since P is more volatile than Nb; as the surface area to volume ratio of the powder is higher than that of the bulk sample, P is lost from the powder first, creating a partial pressure of P in the ampoule. Nevertheless, some P was still lost in the annealing process, leaving Sample 2 as more Nb-rich than Sample 1.

The grain size of each sample was determined using EBSD on an SEM. This confirmed that the annealing did result in grain growth in Sample 2. EBSD maps for both samples are shown in Figure S1.

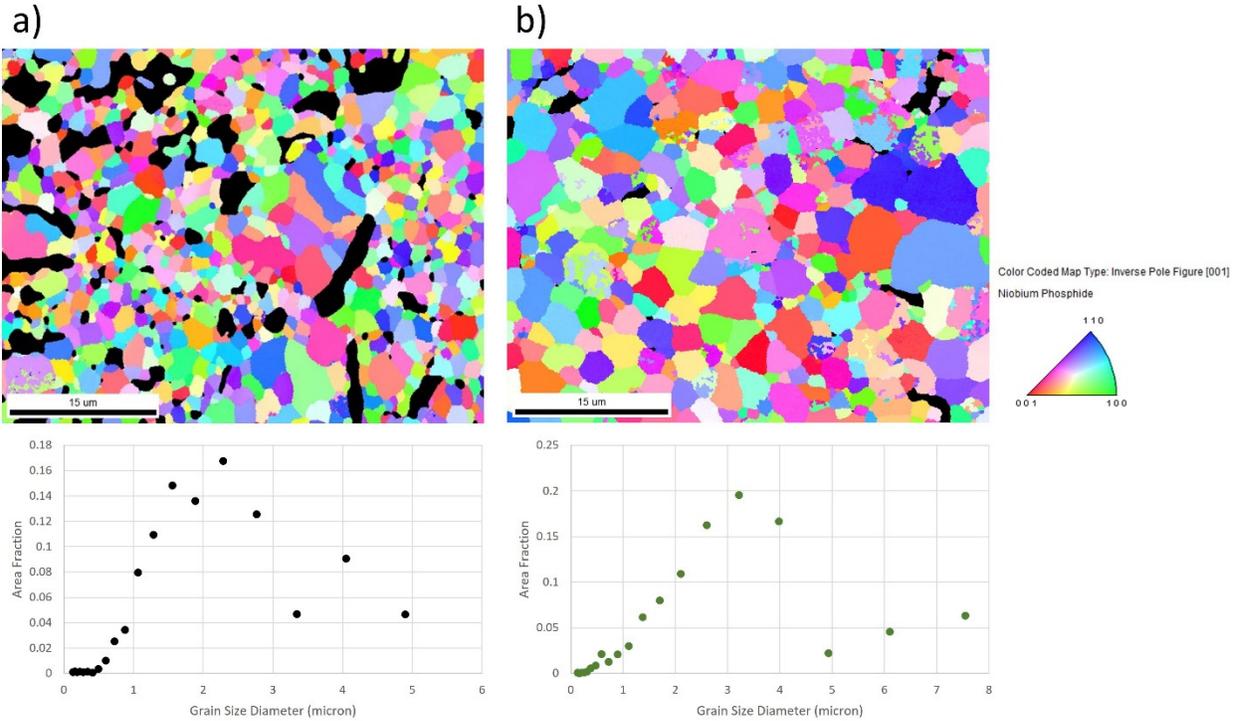

*Figure S1. EBSD results of (a) Sample 1 and (b) Sample 2. Average grain sizes were determined to be 2.18 μm for Sample 1 and 3.08 μm for Sample 2.*

EDS was used to determine the stoichiometry of the samples and showed annealing did result in some change in stoichiometry, with a slightly higher percentage of Nb present in Sample 2, as expected. XPS was also used to analyze the surface stoichiometry of the samples. The surface of the sample was slightly etched during the XPS process to remove a small layer of the sample surface of about 5-10 nm. A comparison to the unetched results showed that very little oxidation was present beyond the first 5-10 nm of both samples.

**II. Experimental Methods**

All transport measurements were completed on a Quantum Design DynaCool Physical Property Measurement System (PPMS). The thermal conductivity, electrical resistivity, Seebeck effect, magneto-Seebeck effect, and Nernst effect were measured using the Thermal Transport Option (TTO) in the conventional one-heater two-thermometer configuration as shown in Figure S2.

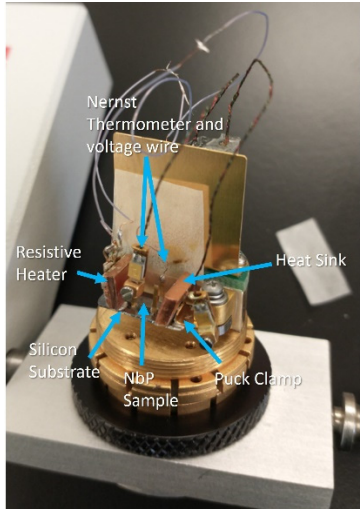

*Figure S2. Example of NbP sample mounted on the TTO puck for transverse transport measurements (note that only one thermometry assembly is attached to the lead in this figure).*

TTO thermometry assemblies were attached to gold-plated copper leads, which were attached to the sample using silver epoxy. The thermometry assemblies include Cernox thermometer sensors and copper voltage wires. Instead of the typical TTO heater assembly, our home-built heater assembly used two 120 Ω strain gauges wired electrically in series and attached to a piece of copper foil to make a resistive heater; this heater assembly was used for thermoelectric transport measurements.

Electrical resistivity and thermal conductivity were measured using Multivu controls, Quantum Design's proprietary software, using the TTO heater assembly. The Seebeck effect, magneto-Seebeck effect, and Nernst effect were measured with the home-built resistive heater assembly described above and using a home-built electrical breakout system with precision external electronics and customized controls code written in Labview. The Hall effect was measured using the PPMS's Electrical Transport Option (ETO) using the van der Pauw method, also controlled via Multivu.

The Seebeck, magneto-Seebeck, and Nernst effects were measured both adiabatically and isothermally, however only the isothermal results are presented in the main text. Previous adiabatic measurements of an NbP single crystal showed that the adiabatic magneto-Seebeck thermopower ($S_{xxz}$) measured mapped to the isothermal Nernst thermopower ($S_{xyz}$) at similar temperatures and magnetic fields [1]. This indicated that despite traditional longitudinal setup, a transverse thermopower was measured due to parasitic contributions from the Nernst effect via the large thermal Hall effect present in NbP (and other Weyl semimetals). The effects of this large thermal Hall effect on the measurement setup are illustrated Figure S3, showing that the isothermal mount maintains isotherms in the y-direction, while an adiabatic mount does not; therefore, an adiabatic sample mount allows parasitic contributions to be measured in both longitudinal and transverse thermoelectric transport.

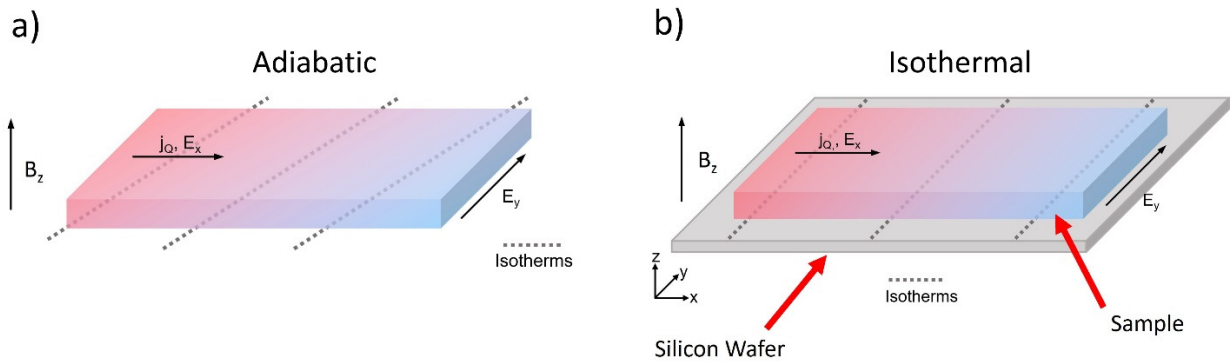

*Figure S3. Adiabatic sample mount (a) compared to the isothermal mount (b). The silicon substrate in the isothermal mount ensures isotherms parallel to the y-axis of the figure, whereas heat flows unrestricted and subject to the thermal Hall effect in the adiabatic mount. Reproduced from Ref. [2], with the permission of AIP Publishing.*

Figure S3(a) shows the behaviors of a sample mounted adiabatically, experiencing a large thermal Hall effect. The solution to this is to mount the sample isothermally rather than adiabatically. For the isothermal mount shown in Figure S3(b), the sample is attached to a silicon wafer using GE varnish so that the two are in thermal contact but electrically insulated from each other. The silicon experiences no thermal Hall effect and has a thermal conductivity much larger than the NbP sample, so it effectively controls the temperature gradient and ensures isotherms in the *y*-direction. The heater is also in thermal contact with the silicon, so the heat flux is driven through the silicon wafer. A comparison of the adiabatic mounting technique in comparison to an isothermally prepared sample is shown in Figure S4.

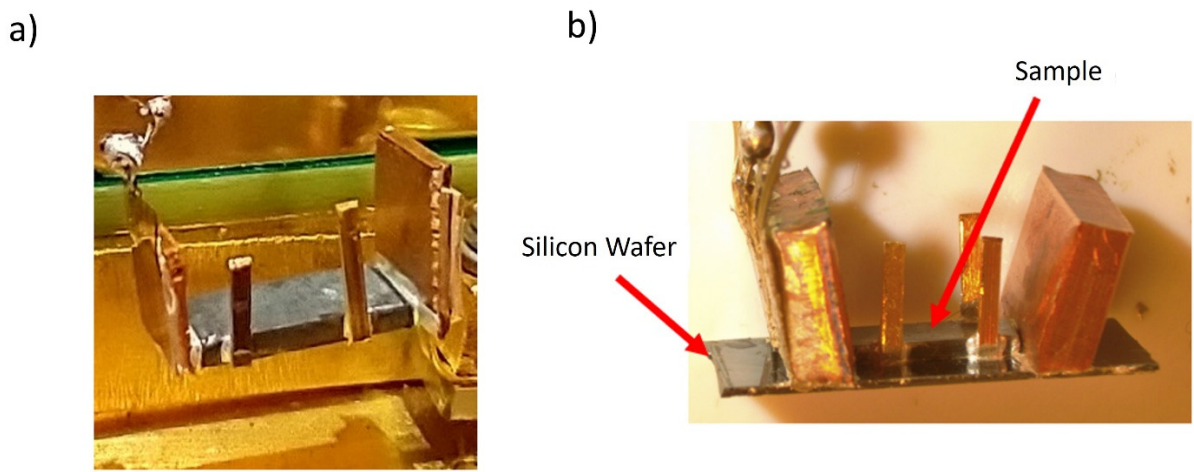

*Figure S4. Sample Preparation for: (a) adiabatic measurement and (b) isothermal measurement.*

The isothermal mount, Figure S4(b), is near identical to the adiabatic preparation, Figure S4(a), with the only difference being the addition of the silicon wafer. The electrical resistivity and thermal conductivity were measured using the adiabatic mount. The Seebeck and Nernst effects were measured isothermally, and the magneto-Seebeck effect was measured both adiabatically and isothermally to confirm its existence since no isothermal magneto-Seebeck effect, but an adiabatic magneto-Seebeck effect, was observed in the single-crystalline sample of Ref. [1].

An overhead view of an isothermally prepared sample of NbP is shown in Figure S5.

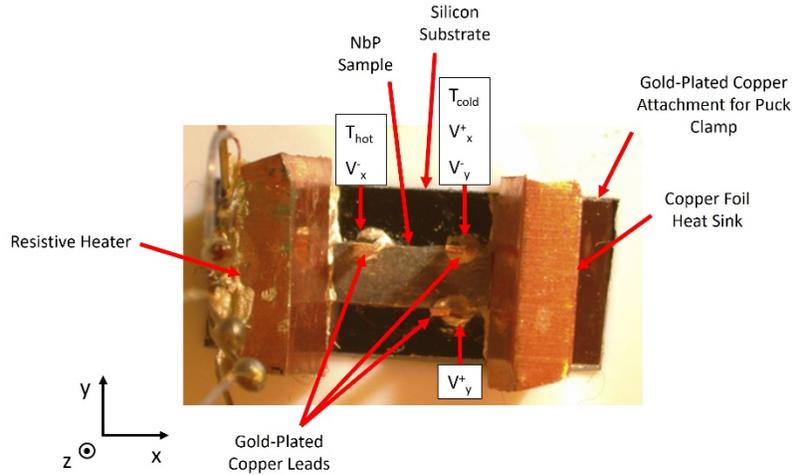

*Figure S5. Overhead view of the NbP Sample 1 mounted for isothermal TTO measurement of both Nernst and Seebeck effects.*

Gold-plated copper leads were attached using silver epoxy both longitudinally and transversely. The direction of the voltage and temperature difference measurements correlating to the leads are shown using the figure axis. From this view, any applied magnetic field would be coming out of the page. Two pieces of copper foil were attached on either end of the sample. On the left end of the sample, two resistive heaters were attached in series and epoxied to the copper foil. On the right end of the sample, the right-hand piece of copper foils acts as a heat sink.

## III. Experimental Results

Figure S6 shows the difference between adiabatic and isothermal $S_{xxz}$ at 9 T. The adiabatic data is consistently smaller than isothermal data due to the presence of a parasitic Nernst effect via the thermal Hall effect. The Nernst thermopower, $S_{xyz}$, of the NbP samples is positive in a positive applied magnetic field causing the measured adiabatic $S_{xxz}$ to be significantly smaller than isothermal $S_{xxz}$ due to the positive $S_{xyz}$ counteracting the negative $S_{xxz}$.

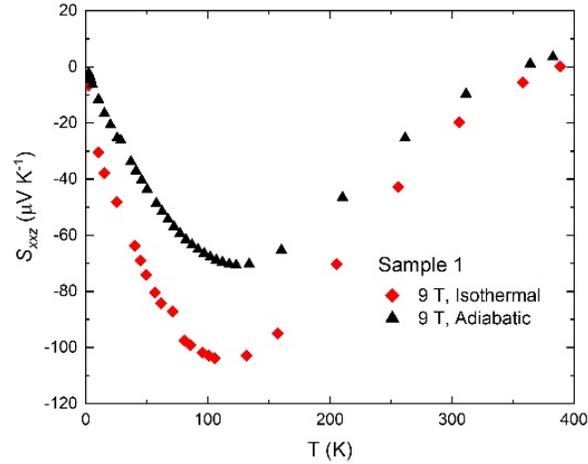

*Figure S6. Comparison of adiabatic and isothermal Seebeck thermopower as functions of temperature in Sample 1*

This adiabatic Seebeck data was measured in Sample 1 to confirm that the isothermal magneto-Seebeck effect is in fact real and not the result of improper mounting or measurement techniques. With this confirmation, the remaining thermoelectric measurements were conducted isothermally.

Hall measurements were completed on both samples using Quantum Design's ETO (electrical transport option) in the Van der Pauw configuration. The Hall resistivity, $\rho_{xyz}$, is shown for Sample 1 and Sample 2 in Figure S7 and Figure S8 respectively.

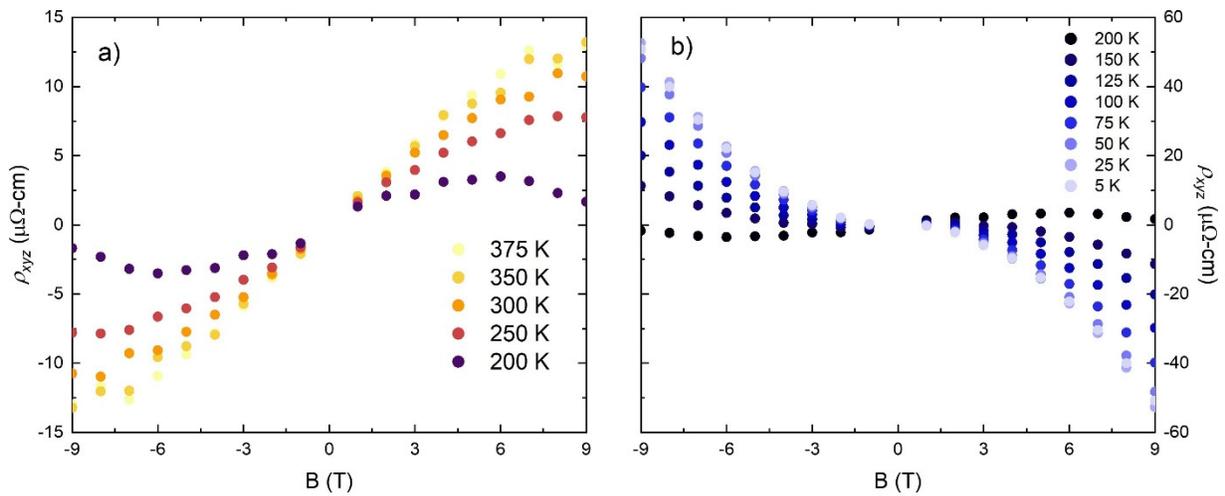

*Figure S7. Magnetic field dependence of the Hall resistivity fir Sample 1 at (a) high temperature and (b) low temperature.*

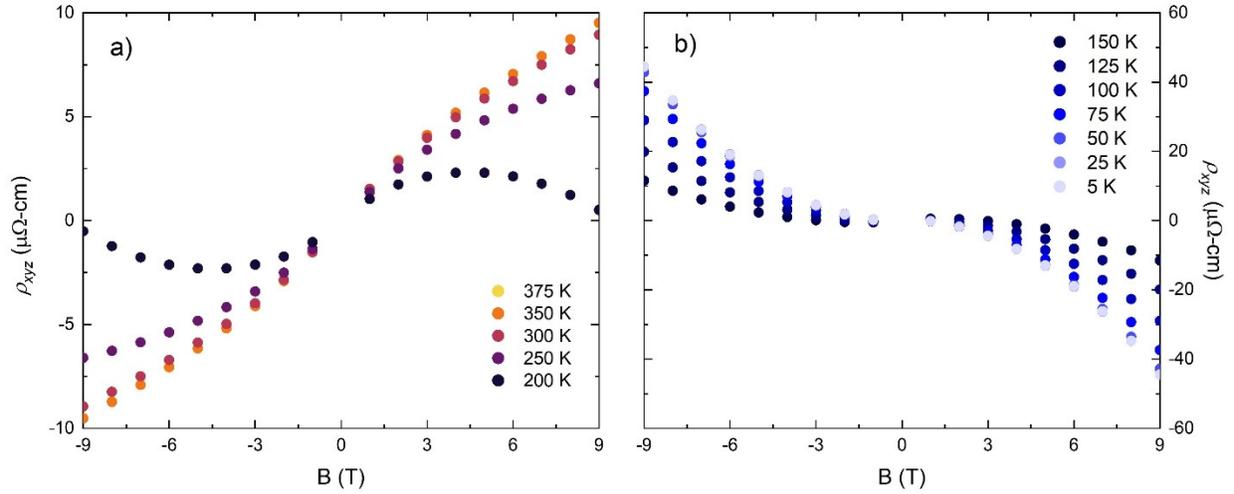

*Figure S8. Magnetic field dependence of the Hall resistivity for Sample 2 at (a) high temperature and (b) low temperature*

Data from both samples indicates a positive Hall coefficient above ~200 K and a negative Hall coefficient at lower temperatures. Both samples then exhibit a transition from positive to negative Hall coefficient between ~150 – 200 K.

      The electron density in the 0 K limit was then estimated by fitting the Hall resistivity of Samples 1 and 2 to a two-carrier model. This same process was repeated using Hall resistivity data taken from Ref. [3] for single-crystalline NbP and Ref. [4] for the previously studied results on polycrystalline NbP. The resulting electron densities are summarized in Table SI. These results indicate that the polycrystalline samples have smaller electron densities than the single-crystalline sample, which is consistent with the theoretical results in the main text.

*Table SI. Estimated charge carrier density for single-crystalline and polycrystalline NbP*

| Source | Electron Density in the 0 K limit [$cm^{-3}$] |
|---|---|
| Single-crystalline NbP from Ref. [3] | $1.96 \times 10^{18}$ |
| Polycrystalline NbP from Ref. [4] | $9.00 \times 10^{17}$ |
| Polycrystalline NbP, Sample 1 [this work] | $7.65 \times 10^{17}$ |
| Polycrystalline NbP, Sample 2 [this work] | $4.00 \times 10^{17}$ |

      Figure S9 shows the isothermal magneto-Seebeck thermopower, $S_{xxz}$, as a function of magnetic field. Both samples exhibit a linear relationship between $S_{xxz}$ and applied magnetic field, which is consistent with theoretical models.

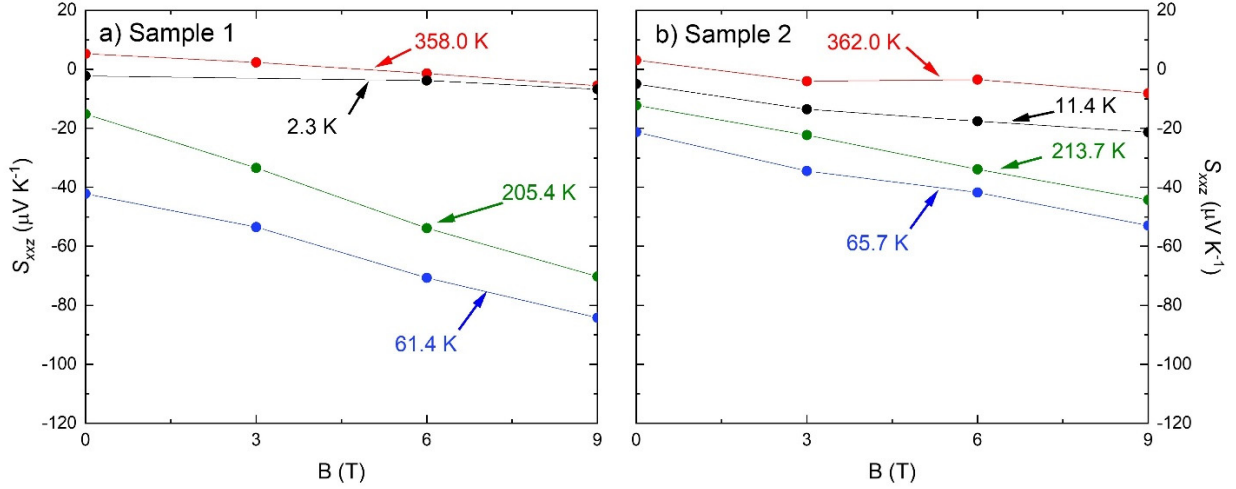

*Figure S9. Magnetic field dependence of $S_{xxz}$ at select temperatures for (a) Sample 1 and (b) Sample 2.*

**IV. Theoretical Results**

    Here we describe our theoretical calculations of the Seebeck and Nernst thermopowers, $S_{xxz}$ and $S_{xyz}$ respectively, as functions of doping, temperature, and magnetic field. We assume everywhere that the thermoelectric transport is dominated by a Weyl cone with degeneracy $g$ (spin + valley), and for simplicity we take an isotropic Weyl velocity $v_F$.

    In general, the transport of heat and electrical current in the presence of a temperature gradient $\vec{\nabla}T$ and electric field $\vec{E}$ are defined by the Equations (S1) and (S2):

$$\vec{J^E} = \hat{\sigma}\vec{E} - \hat{\alpha}\vec{\nabla}T \tag{S1}$$

$$\vec{J^Q} = T\hat{\alpha}\vec{E} - \hat{\kappa}\vec{\nabla}T \tag{S2}$$

Here, $\vec{J^E}$ and $\vec{J^Q}$ are the electrical and heat current densities, respectively. $\hat{\sigma}$ denotes the electrical conductivity tensor, $\hat{\kappa}$ is the thermal conductivity tensor, and $\hat{\alpha}$ is the Peltier conductivity tensor. The thermoelectric tensor $\hat{S}$ is given by $\hat{S} = \hat{\sigma}^{-1}\hat{\alpha}$, so that the Seebeck and Nernst thermopowers are:

$$S_{xxz} = \frac{\alpha_{xxz}\sigma_{xxz} + \alpha_{xyz}\sigma_{xyz}}{\sigma_{xxz}^2 + \sigma_{xyz}^2} \tag{S3}$$

$$S_{xyz} = \frac{\alpha_{xyz}\sigma_{xxz} - \alpha_{xxz}\sigma_{xyz}}{\sigma_{xxz}^2 + \sigma_{xyz}^2} \tag{S4}$$

    In the semiclassical limit, where the magnetic field is weak enough that the cyclotron energy is small compared to the typical electron kinetic energy, the electrical conductivity tensor $\hat{\sigma}$ and the Peltier

conductivity tensor $\hat{\alpha}$ can be calculated using a semiclassical Boltzmann transport description. In this case [5,6,7]:

$$\sigma_{xxz(xyz)} = \int d\epsilon \left(-\frac{\partial f}{\partial \epsilon}\right) \sigma_{xxz(xyz)}(\epsilon) \tag{S5}$$

$$\alpha_{xxz(xyz)} = \frac{1}{eT} \int d\epsilon (\epsilon - \mu) \left(-\frac{\partial f}{\partial \epsilon}\right) \sigma_{xxz(xyz)}(\epsilon) \tag{S6}$$

where these equations hold for either the longitudinal (xxz) or transverse (xyz) orientation as indicated by the subscript indices on both $\sigma$ and $\alpha$. Here, $\epsilon$ denotes the quasiparticle energy, $f(\epsilon) = \left[\exp\left(\frac{\epsilon-\mu}{k_B T}\right) + 1\right]^{-1}$ is the Fermi function, with $\mu$ the chemical potential, and e is the fundamental charge. The quantities $\sigma_{xxz(xyz)}(\epsilon)$ denote the zero-temperature conductivity for a system with chemical potential equal to $\epsilon$. Within the scattering-time approximation these quantities are given by Equations (S7) and (S8):

$$\sigma_{xxz}(\epsilon) = \frac{1}{3} \frac{e^2 \nu(\epsilon) v_F^2(\epsilon) \tau(\epsilon)}{1 + \omega_c^2(\epsilon) \tau^2(\epsilon)} \tag{S7}$$

$$\sigma_{xyz}(\epsilon) = \frac{1}{3} \frac{e^2 \nu(\epsilon) v_F^2(\epsilon) \tau(\epsilon)}{1 + \omega_c^2(\epsilon) \tau^2(\epsilon)} \omega_c(\epsilon) \tau(\epsilon) \tag{S8}$$

Here, $\omega_c(\epsilon) = eBv_F^2/\epsilon$ is the cyclotron frequency, which for linearly-dispersing electrons is energy-dependent. $\tau(\epsilon)$ denotes the transport scattering time, which in general is a function of the quasiparticle energy, and $\nu(\epsilon)$ is the density of states for a Weyl dispersion:

$$\nu(\epsilon) = \frac{g}{2\pi^2} \frac{\epsilon^2}{\hbar^3 v_F^3} \tag{S9}$$

In the main text, we focus on the case where the transport scattering time $\tau$ is an energy-independent constant. In general, $\tau$ may depend on energy in a power-law way, $\tau(\epsilon) \propto \epsilon^\gamma$, where the exponent $\gamma$ depends on the dominant scattering mechanism -- for example, $\gamma = 0$ for short-range scatterers and $\gamma = 4$ for Coulomb impurity scattering [8].

In the limit of zero temperature, the chemical potential $\mu$ (defined relative to the Weyl point) is set by the electron concentration $n$ and is equal to the Fermi energy $E_F$, such that $\mu = E_F = \hbar v_F (6\pi^2 n/g)^{1/3}$. When the temperature is comparable to or greater than the Fermi temperature $T_F = E_F/k_B$, the chemical potential is determined from the charge neutrality condition, which states that the difference between the number of conduction band electrons and the number of valence band holes is fixed:

$$n = \int_0^\infty d\epsilon\, \nu(\epsilon) f(\epsilon) - \int_{-\infty}^0 d\epsilon\, \nu(\epsilon)(1 - f(\epsilon)) \tag{S10}$$

$$n = \frac{g}{6\pi^2} \frac{\mu(\mu^2 + \pi^2 (k_B T)^2)}{\hbar^3 v_F^3} \tag{S11}$$

Solving this equation for $\mu$ gives the value of the chemical potential at a given temperature.

In general, evaluating Equations (S1) through (S4) yields the values of $S_{xxz}$ and $S_{xyz}$ at a given value of $T$, $B$, and $n$. Such calculations take as an input the Fermi energy $E_F$, the Fermi velocity $v_F$, and the scattering time $\tau$. The theory curves shown in the main text are produced by a numeric evaluation of Equations (S1) through (S4), with the listed values of the three parameters. In the low temperature limit, $T \ll T_F$, Equations (S1) through (S4) are equivalent to the more common Mott formula for the thermoelectric tensor:

$$\hat{S} = -\frac{\pi^2}{3} k_B^2 T \hat{\sigma}^{-1} \frac{d\hat{\sigma}}{dE} \tag{S12}$$

which can be shown by applying the Sommerfeld expansion to Equations (S1) through (S4).

# References

[1] S. J. Watzman, T. M. McCormick, C. Shekhar, S. C. Wu, Y. Sun, A. Prakash, C. Felser, N. Trivedi, and J. P. Heremans. Dirac dispersion generates unusually large Nernst effect in Weyl semimetals. *Phys. Rev. B* **97**, 161404(R) (2018).

[2] S. R. Boona, H. Jin, and S. J. Watzman. Transverse thermal energy conversion using spin and topological structures. *J. Appl. Phys*. **130**, 171101 (2021).

[3] C. Shekhar, A. K. Nayak, Y. Sun, M. Schmidt, M. Nicklas, I. Leermakers, U. Zeitler, Y. Skourski, J. Wosnitza, Z. Liu, Y. Chen, W. Schnelle, H. Borrmann, Y. Grin, C. Felser, and B. Yan. Extremely large magnetoresistance and ultrahigh mobility in the topological Weyl semimetal candidate NbP. *Nat. Phys*. **11** (2015).

[4] C. Fu, S. N. Guin, S. J. Watzman, G. Li, E. Liu, N. Kumar, V. Su, W. Schnelle, G. Auffermann, C. Shekhar, Y. Sun, J. Gooth, and C. Felser. Large Nernst power factor over a broad temperature range in polycrystalline Weyl semimetal NbP. *Energy Environ. Sci*. **11**, 2813 (2018).

[5] N. W. Ashcroft and N. D. Mermin. *Solid State Physics*. New York: Holt, Rinehart, and Winston (1976).

[6] V. Kozii, B. Skinner, and L. Fu. Thermoelectric Hall conductivity and figure of merit in Dirac/Weyl materials. *Phys. Rev. B* **99,** 155123 (2019).

[7] X. Feng and B. Skinner. Large enhancement of thermopower at low magnetic field in compensated semimetals. *Phys. Rev. Materials* **5**, 024202 (2021).

[8] B. Skinner. Coulomb disorder in three-dimensional Dirac systems. *Phys. Rev. B* **90**, 060202(R), (2014).